\documentclass[useAMS,usenatbib,usedcolumn]{mn2e}

\usepackage{graphicx,color,soul} 
\usepackage{times}

\title[Observations of two GPS pulsars]{Multi-frequency observations and spectral analysis of two gigahertz-peaked spectra pulsars}
{ }
\author[Ro\.{z}ko et al.]{K. Ro\.{z}ko$^{1}$\thanks{e-mail: krozko@gmail.com}, K. M. Rajwade$^{2,3,4}$, W. Lewandowski$^{1}$, R. Basu$^{1,5}$, J. Kijak$^{1}$ and D. R. Lorimer$^{2,3}$\\
$^{1}$Janusz Gil Institute of Astronomy, University of Zielona G\'ora, Zielona G\'ora, Poland
\\$^{2}$Department of Physics and Astronomy, West Virginia University, Morgantown, WV, 26506 USA
\\$^{3}$ Centre for Gravitational Physics and Cosmology, West Virginia University, Chestnut Ridge Research building, Morgantown, WV 26505, USA
\\$^{4}$Jodrell bank Centre for Astrophysics, University of Manchester, Oxford Road, Manchester M193PL, UK
\\$^{5}$Inter-University Centre for Astronomy and Astrophysics, Pune, 411007, India}

\begin{document}

\date{Accepted\ldots Received\ldots ; in original form\ldots}


\maketitle

\label{firstpage}

\begin{abstract}
We report the multi-frequency observations of two pulsars: J1740$+$1000 and B1800$-$21, using the Giant Metrewave Radio Telescope and the Green Bank Telescope. The  main aim of these observations was to estimate the flux density spectrum of these pulsars, as both of them were previously reported to exhibit gigahertz-peaked spectra. J1740$+$1000 is a young pulsar far from the Galactic plane and the interpretation of its spectrum was inconclusive in the light of the recent flux density measurements. Our result supports the gigahertz-peaked interpretation of the PSR J1740$+$1000 spectrum. B1800$-$21 is a Vela-like pulsar near the W30 complex, whose spectrum exhibit a significant change between 2012 and 2014 year. Our analysis shows that the 
current shape of the spectrum is similar to that observed before 2009 and confirms that the observed spectral change happen in a time-scale of a few years. 

 \end{abstract}

\begin{keywords}
pulsars: general -- pulsars: individual: J1740$+$1000, B1800$-$21 (J1803$-$2137)
\end{keywords}

\section{Introduction}
Radio flux density is one of the pulsar's main observables. The flux density spectrum provides information about pulsar intrinsic emission properties, and since the radiation is affected by the interstellar medium (ISM), it can also provide information about the propagation effects. Fifty years have passed since the discovery of pulsars \citep{h68} and to date we know more than 2600 of these objects (see the ATNF Pulsar Catalogue\footnote{http://www.atnf.csiro.au/people/pulsar/psrcat/}, 
\citealt{man2005}). However, in most cases their flux density was measured only in a~limited radio frequency range and therefore we know nothing about the spectra of such pulsars. Currently, we have flux density measurements at multiple radio frequencies for only a small subset of pulsars \citep[less than 500, see ][]{lorimer1995,Maron2000,Kramer1998,Jankowski2018}. 

Most of the pulsar spectra can be described by a single power-law function with the population average spectral index close to $-1.6$ \citep{lorimer1995,Jankowski2018}. In addition, from simulations of the pulsar population, \citet{bates2013} concluded that the underlying population of pulsars is described, to first order, by a single power law spectrum with a~Gaussian distribution of mean $-1.4$ and standard deviation of $0.9$. However, spectra of some pulsars show different behaviour. Among them are the gigahertz-peaked spectra (GPS) pulsars characterized by a positive spectral index in the frequency range below the peak frequency. These turnovers peak at frequencies around one gigahertz (see \citealt{kijak2011a,kijak2011b,kijak2017}). The number of GPS pulsars remains small and their origin is not fully understood. One hypothesis proposes that this turnover is caused by the free-free thermal absorption in the medium surrounding the pulsar or in the general ISM \citep{sie73}. The role of such absorbers may be fulfilled by dense supernova remnant filaments, the pulsar wind nebulae or HII regions (see \citealt{lewandowski2015a} and \citealt{rajwade2016}). In the free-free absorption model the flux at low frequencies is heavily absorbed while at higher frequencies these objects exhibit a typical power-law spectrum, which is basically an unchanged intrinsic pulsar spectrum. Therefore observations at multiple radio frequencies are necessary to model the pulsar spectra using free-free absorption. The other possible mechanism that could cause a turnover in the pulsar spectrum is the synchrotron self-absorption, that was proposed by various authors to explain turnovers at lower frequencies ($\sim 100$~MHz, see e.g. \citealt{sie73,i98}). GPS pulsars were not discovered until 2007 \citep{kijak2007} and it is very difficult to explain high frequency turnovers with synchrotron self-absorption. Obviously, in the case of the interferometric flux density measurements we cannot be sure that the observed total flux comes only from the coherent mechanism: in general it could be sum of the pulsed and unpulsed emission. Nevertheless we believe that the synchrotron self-absorption is a very unlikely cause for the pulsar spectra turnovers.

In this paper we investigate the spectra of two peculiar pulsars: J1740$+$1000 and B1800$-$21. J1740$+$1000 is a young pulsar far from the galactic plane. B1800$-$21 is a Vela-like pulsar located near the W30 complex. Both pulsars are associated with the pulsar wind nebulae (PWNe). Recently, \citet{Basu2018} have shown that several pulsars with PWNe have a GPS. In the case of PSR~J1740$+$1000 the previous flux measurements were highly spread out (see Figure~\ref{1740f2}), which makes the correct interpretation of spectral behaviour very difficult. Based on their measurements \citet{kijak2011b} described the spectrum of PSR~J1740$+$1000 using a simple power-law function. \citet{dembska2014} added three measurements obtained using the Effelsberg Telescope (at frequencies of 2600~MHz,~4850~MHz and 8350~MHz). The authors decided also to exclude the flux density measurement at 1070~MHz from the fitting procedure, because it was substantially smaller than values obtained at neighbouring frequencies probably due to interstellar scintillations. That set of data points allowed them to classify the spectrum of PSR~J1740$+$1000 as a GPS. That interpretation was further strengthened by an unpublished interferometric measurement at 325~MHz from the Giant Meterwave Radio Telescope (observed in January, 2015). However, based on the subsequent flux density measurement at 150~MHz using the LOFAR telescope, \citet{bilous2016} suggested a simple power-law interpretation of the spectrum. 

 In the case of PSR~B1800$-$21 we observed that its spectrum changed shape over time-scale of a few years (see \citealt{basu2016}, and references therein), which prompted us to continue our investigation. We would like to point out that in the case when a pulsar spectrum is variable (with a time-scale of a few years in this case) using the flux measurements from very different epochs to reconstruct it may lead to confusing results. Moreover, for both pulsars there were no recent measurements at high frequencies, which are crucial for constraining the intrinsic pulsar spectral index using the free-free absorption model. 

We have conducted an observational campaign to measure flux density for both pulsars over a wide frequency range: from 325~MHz to 5900~MHz. The observations were carried out over a short period of time, between August 2016 and January 2017, to minimize the influence of a potential spectral change. We used the Giant Meterwave Radio Telescope at 325~MHz, 610~MHz, 1280~MHz and the Green Bank Telescope at 900~MHz, 1600~MHz, 2150~MHz and 5900~MHz. We also used the free-free thermal absorption model to characterize the spectral nature. In the case of PSR~J1740$+$1000 the recent result support the GPS interpretation of its spectrum. In the case of PSR~B1800$-$21 we show that its current peak frequency is equal 760~MHz, which is similar to the peak frequency of the thermal absorption model fitted to data measured before 2009 \citep{basu2016}. 

The outline of the paper is as follows. In Section 2 we describe the observations and present the methods used in our analysis. Section 3 details the results of spectral analysis, and in Section 4 we discuss the implications of our analysis on the interpretation of the spectra of PSRs J1740+1000 and B1800-21. Our findings are summarized in Section 5.

\section{Observations and data analysis}

The first part of our project involved observations of the two pulsars J1740+1000 and B1800-21 using the Giant Meterwave Radio Telescope (GMRT) located near Pune in India \citep{Swarup1991}. The GMRT array consists of 14 central antennas located within a~square kilometer region and an additional 16 antennas spread out along the three arms. Each antenna has 45-m diameter and is fully steerable. The GMRT receiver system allows to observe pulsars simultaneously in two modes: interferometric and phased array (see for details~\citealt{basu2016}).

We observed both pulsars at three frequencies: 325~MHz, 610~MHz and 1200~MHz. Each of them during three observational sessions at each frequency separated by at least one week to account for the possible influence of interstellar scintillations (for a~summary of that phenomenon see \citealt{rickett1990} and \citealt{gupta1995}).
Observational epochs are included in Table~\ref{tabobs}. The  33~MHz bandwidth at each frequency band was divided into 256 channels. Both pulsars were observed for around 35-minutes during each session (the only exception was August~1 when we observed for twice as long). We used standard observational procedure for our analysis. We recorded the flux calibrators 3C286 and 3C48 (the latter only for the September 24, 2016 observations) at the beginning or at the end of the observing session. These are standard flux calibrators \citep{perley2017}. For the phase calibrators we have chosen stable, point-like sources which were less than 20 degrees away from our targets. The sources were selected from the list of VLA Calibrators\footnote{https://science.nrao.edu/facilities/vla/observing/callist}.
Selected phase calibrators were interspersed at regular intervals to correct for the temporal variations in the antenna gains. These are listed in Table~\ref{tabobs}. 

The editing out of spurious data, calibration, both across the frequency band and for temporal variations, and imaging were carried out using the Astronomical Image Processing System as previously described by \citet{dembska2015b}. We set the flux scales of the calibrators 3C286 and 3C48 using the latest \citet{perley2013} estimates which were used to calculate the flux values for phase calibrators (see Table~\ref{tabobs}). 

In the phased array mode signals from the all available central antennas, and two or three nearest arm antennas (around 20 in total) were co-added producing time series data with resolutions of 122~$\mu$s. To increase the signal-to-noise ratio the antennas were phase aligned using the phase calibrators before recording the source. Data were collected using the GMRT Software Backend \citep[GSB,][]{roy2010}. We observed all the calibrators using the position switching method to estimate the flux levels of the phased-array i.e. data were recorded away from source before and after the source pointing. The data was dedispersed and folded using the topocentric pulsar period. The  obtained integrated profiles are presented in the Appendix (see Figures~\ref{ap1} and \ref{ap2}). The 'ON-OFF' levels for the calibrators showed significant changes due to temporal variations in the receiver system. Hence, we were not able to constrain the pulsar flux density accurately using the phased-array mode. These temporal instabilities were only an issue for the phased-array mode observations where the signal from individual antennas were recorded. In the interferometric mode the signals recorded are correlations between antenna pairs which resolve out any variations in individual antennas. This makes the interferometric mode much more precise and stable for flux density measurements. However, as \citet{basu2016} showed previously, the interferometric method and the phased array method give comparable results. 

The Robert C. Byrd Green Bank Telescope (GBT) is the world's largest, fully steerable, single dish telescope located at Green Bank, West Virginia, USA \citep{oneil2008}.
The GBT observations were carried out in the 16A semester (MJD 57601--57764). The pulsars 
were observed across 800~MHz of bandwidth using Green Bank Ultimate Pulsar Processing Instrument \citep[GUPPI;~][]{ran2009} as the 
pulsar backend in fold mode. The observations were conducted at four frequency bands centered at the following frequencies: 900~MHz, 1600~MHz, 2150~MHz and 5900~MHz.
We observed lower frequencies at multiple epochs, that 
were separated by a few days to account for interstellar scintillation. For each pulsar,
 the data were coherently dedispersed at the pulsar DM and folded at the topocentric 
 period to generate a folded pulse profile. Each sub-integration of the folded 
 profile, spanning 10~seconds was saved to disk. Before the start of each observation, 
 we observed a standard flux calibrator for two minutes to calibrate the data 
 offline. The details of the GBT observations are given in Table~\ref{tabobsGBT}. Unfortunately, we had problems during our calibrator observations that rendered those data unusable for flux calibration. Hence, we had to use calibrator observations that are regularly done as part of observations for the North American Nanohertz Observatory for Gravitational Waves~\citep{mcl2013}. The calibrator observations as part of this project and our observations are separated by a day or so, hence, we assumed that the instrument noise characteristics do not change over that time and used those data for calibrating our datasets. There were no calibrator observations available for the 5.9~GHz observations, and therefore we used the radiometer equation~\citep[see, e.g.,][]{lor2004} to flux calibrate the data. This explains the larger relative error on the flux values (see Table~\ref{tabFlux2}). Example profiles for both pulsars obtained from the GBT observations are presented in Fig.~\ref{p_ex}. The full set of profiles is presented in the Appendix (see Figures~\ref{ap1} and \ref{ap2}).

\begin{table}
\resizebox{\hsize}{!}{
\begin{minipage}{80mm}
\caption{Observing details for the GMRT observations.}
\centering
\begin{tabular}{cc D{,}{\pm}{3.3}}
\hline
 & & \\
Obs Date & Phase Calibrator & \multicolumn{1}{c}{Calibrator Flux}\\
 &  & \multicolumn{1}{c}{{\footnotesize Jy}}\\
\hline
\multicolumn{3}{c}{325~MHz}  \\
 01 Aug, 2016 & 1822$-$096 & 3.8, 0.2 \\
 14 Aug, 2016 & 1822$-$096 & 3.6, 0.2 \\
 27 Aug, 2016 & 1822$-$096 & 3.7, 0.2\\
 & & \\
\multicolumn{3}{c}{610~MHz}  \\
 15 Aug, 2016 & 1822$-$096 &  6.2, 0.4 \\
 29 Aug, 2016 & 1822$-$096 &  6.5, 0.4 \\
 24 Sep, 2016 & 1822$-$096 &  6.7, 0.5\\
 & & \\
\multicolumn{3}{c}{1280~MHz}  \\
 08 Aug, 2016 & 1743$-$038 & 2.3, 0.2 \\
 08 Aug, 2016 & 1714$-$252 & 2.6, 0.2 \\
 23 Aug, 2016 & 1743$-$038 & 2.1, 0.1  \\
 23 Aug, 2016 & 1714$-$252 & 2.4, 0.2 \\
 03 Sep, 2016 & 1743$-$038 & 2.5, 0.2 \\
 03 Sep, 2016 & 1714$-$252 & 2.7, 0.2 \\
 
 & & \\
\hline
\end{tabular}
\label{tabobs}
\end{minipage}
}
\end{table}

\begin{table}
\resizebox{\hsize}{!}{
\begin{minipage}{80mm}
\caption{Observing details for the GBT observations.}
\centering
\begin{tabular}{cccc}
\hline
 & & & \\
Frequency & \multicolumn{3}{c}{Obs date}\\
\multicolumn{1}{c}{{\footnotesize MHz}} &  & & \\
\hline
\multicolumn{4}{c}{J1740$+$1000}  \\
900 & 12 Dec, 2016 & 14 Dec, 2016 &  \\
1600 & 25 Nov, 2016 & 8 Dec, 2016 & 22 Dec, 2016 \\
2150 & 23 Dec, 2016 &  & \\
5900 & 27 Nov, 2016 & 24 Dec, 2016 & 19 Jan, 2017 \\
\multicolumn{4}{c}{B1800$-$21}  \\
900 & 01 Jan, 2017 &  &  \\
1600 & 16 Dec, 2016 & 18 Dec, 2016 &  \\
2150 & 15 Jan, 2017 &  & \\
5900 & 17 Jan, 2017 & & \\
 
 & & \\
\hline
\end{tabular}
\label{tabobsGBT}
\end{minipage}
}
\end{table}

\begin{figure}
\resizebox{\hsize}{!}{\includegraphics{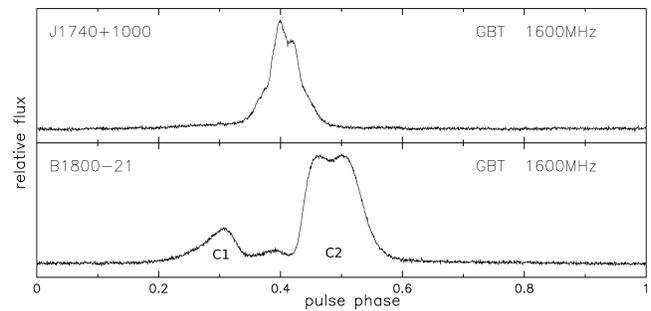}}
\caption{Sample 1600~MHz pulse profiles for both of the pulsars obtained from the GBT.
\label{p_ex}}
\end{figure}

\section{Results and analysis}

Table~\ref{tabFlux1} shows the flux density values determined from the analysis of both, the GMRT data and the GBT data. This table presents flux measurements obtained from individual sessions, as well as their weighted average value. Since the 325~MHz mean flux density values are smaller than the values at 610~MHz, it is clear for both pulsars that their spectral nature deviates from a typical power-law behaviour.   

\subsection{The case of PSR~J1740$+$1000}
  
\begin{table}
\resizebox{\hsize}{!}{
\begin{minipage}{80mm}
\caption{Flux measurements for PSRs~J1740+1000 and B1800$-$21 (J1803$-$2137) for the three separate observations at each frequency ($S_1$, $S_2$ and $S_3$, respectively) and the weighted mean value $\left<S\right>$.}
\centering
\begin{tabular}{@{}l  D{,}{\pm}{3.3}  D{,}{\pm}{3.3} D{,}{\pm}{3.3} D{,}{\pm}{4.4}@{}}
\hline
 & & &  \\
Pulsar & \multicolumn{1}{c}{$S_1$} & \multicolumn{1}{c}{$S_2$} & \multicolumn{1}{c}{$S_3$} & \multicolumn{1}{c}{$\left<S\right>$}\\
& \multicolumn{1}{c}{mJy} & \multicolumn{1}{c}{mJy} &\multicolumn{1}{c}{mJy}&\multicolumn{1}{c}{mJy}\\
\hline

\multicolumn{4}{c}{325 MHz} (GMRT)\\
J1740$+$1000 & 3.4,0.4 & 3.3,0.3 & 4.2,0.6 & 3.5,0.2\\

B1800$-$21 & 3.7,0.3 & 2.7,0.4 & 3.3,0.3 & 3.3,0.3\\
 & & & & \\

\multicolumn{4}{c}{610 MHz} (GMRT)\\

J1740$+$1000 & 5.5,0.4 & 4.3,0.3 & 7.8,0.6 & 5.2,0.8 \\

B1800$-$21 & 9.5,0.7 & 10.4,0.8 & 9.4,1.1 & 9.8,0.5\\

 & & & & \\
 
 \multicolumn{4}{c}{900 MHz} (GBT)\\

J1740$+$1000 & 6.5,0.7 & 7.8,0.8 &  & 6.9,0.7 \\

B1800$-$21 & 21.4,2.1 &  &  & 21.4,2.1\\

 & & & & \\

\multicolumn{4}{c}{1280 MHz} (GMRT)\\

J1740$+$1000 & 1.7,0.2 & 6.2,0.4 & 3.6,0.3 & 2.7,1.1\\

 B1800$-$21& 10.5,1.0 & 7.4,1.1 & 12.0,1.6 & 9.6, 1.3\\
 
  & & & & \\
 
 \multicolumn{4}{c}{1600 MHz} (GBT)\\

J1740$+$1000 & 4.5,0.5 & 2.4,0.2 & 3.1,0.3 & 2.9,0.5\\

B1800$-$21 &  10.3,1.0 & 12.5,1.3 &  & 11.1,1.1\\

 & & & & \\

\multicolumn{4}{c}{2150 MHz} (GBT)\\

J1740$+$1000 & 1.3,0.1 &  &  & 1.3,0.1\\

 B1800$-$21 & 9.3,0.9  & & & 9.3,0.9\\
 
  & & & & \\
  
  \multicolumn{4}{c}{5900 MHz} (GBT)\\

J1740$+$1000 & 0.9,0.3 & 0.5,0.3 & 0.4,0.3 & 0.6,0.2\\

 B1800$-$21 & 2.2,0.3  & & & 2.2,0.3\\
 
  & & & & \\

\hline
 & & &  \\
\end{tabular}
\label{tabFlux1}
\end{minipage}
}
\end{table}

 \begin{figure}
\resizebox{\hsize}{!}{\includegraphics{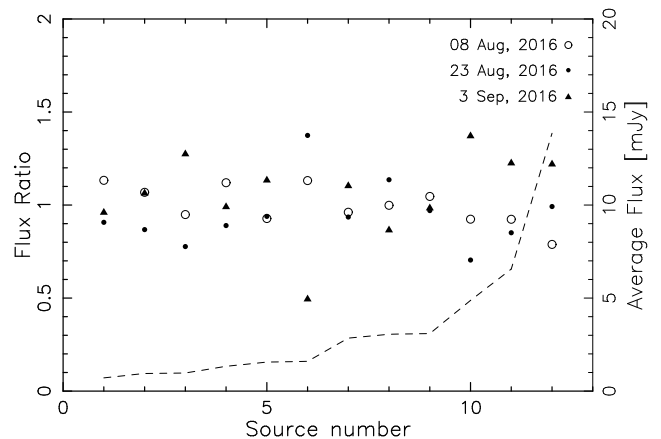}}
\caption{The figure shows the ratio of the flux with respect to the average flux for 12 sources nearby PSR~J1740+1000 across the three observing runs. The standard deviation $\sigma$ equals $0.18$ The dashed line shows the average flux of each source. 
\label{scint1}}
\end{figure}

\begin{table}
\resizebox{\hsize}{!}{
\begin{minipage}{80mm}
\caption{Flux measurements from $\sim$10-minutes scans (obtained from the interferometric GMRT observations at 1280~MHz) for PSR~J1740+1000 and three nearby sources.}
\centering
\begin{tabular}{@{}l  D{,}{\pm}{3.3}  D{,}{\pm}{3.3} D{,}{\pm}{3.3} D{,}{\pm}{4.4}@{}}
\hline
 & & &  \\
Day & \multicolumn{1}{c}{PSR Flux} & \multicolumn{1}{c}{O$_1$ Flux} & \multicolumn{1}{c}{O$_2$ Flux} & \multicolumn{1}{c}{O$_3$ Flux}\\
& \multicolumn{1}{c}{mJy} & \multicolumn{1}{c}{mJy} &\multicolumn{1}{c}{mJy} &\multicolumn{1}{c}{mJy}\\
\hline

8 Aug 1 scan & 1.4,0.3 & 2.9,0.3 & 3.0,0.2 & 10.9,0.3\\

8 Aug 2 scan & 1.4,0.3 & 2.4,0.3 & 3.3,0.1 & 11.2,0.3\\

8 Aug 3 scan & 3.0,0.2 & 2.1,0.3 & 3.1,0.2 & 10.4,0.4\\

\\

23 Aug 1 scan & 4.4,0.2 & 2.3,0.3 & 3.0,0.1 & 12.8,0.3\\

23 Aug 2 scan & 5.0,0.2 & 2.6,0.3 & 2.8,0.1 & 12.9,0.3\\

23 Aug 3 scan & 8.0,0.3 & 2.7,0.3 & 3.0,0.1 & 13.4,0.3\\

\\

3 Sep 1 scan & 3.1,0.2 & 3.0,0.3 & 2.9,0.3 & 16.5,0.3\\

3 Sep 2 scan & 2.7,0.1 & 3.3,0.3 & 3.1,0.2 & 17.2,0.3\\

3 Sep 3 scan & 2.7,0.2 & 3.1,0.3 & 3.2,0.3 & 17.2,0.3\\

3 Sep 4 scan & 4.7,0.2 & 3.2,0.3 & 2.9,0.3 & 16.8,0.3\\

\hline
 & & &  \\
\end{tabular}
\label{tabFlux2}
\end{minipage}
}
\end{table}
 
The flux density values of PSR~J1740$+$1000 at 1280~MHz vary significantly between the different epochs, while at other frequencies they remain constant within the uncertainty levels. We were aware that scintillations could strongly affect this pulsar's flux measurements especially in the frequency range around 1 GHz. Shortly after its discovery, \citet{mcl2002} estimated the diffractive scintillation time-scale for this pulsar:  $t_{\mathrm{DISS}} = 271 \pm 146$ s at~1410~MHz.

To exclude the possibility that the observed flux variations are caused by the receiver system we devised several tests to check its stability at 1280~MHz. First, we measured the flux values for 12 sources in the field of view of PSR~J1740$+$1000 at different epochs to check their variability, similar to \citet{dembska2015b}. We calculated the average flux density value and the ratio between the flux from each observational session with respect to the average flux. Figure~\ref{scint1} summarize our results. We found that the ratios are symmetrically scattered around unity with a small spread ($\sigma=0.18$), which confirms that our analysis was correct and consistent. This makes it unlikely that the instability in the receiver system caused large variations in the PSR~J1740$+$1000 flux measurements. 

\begin{figure*}
\resizebox{\hsize}{!}{\includegraphics[angle=-90]{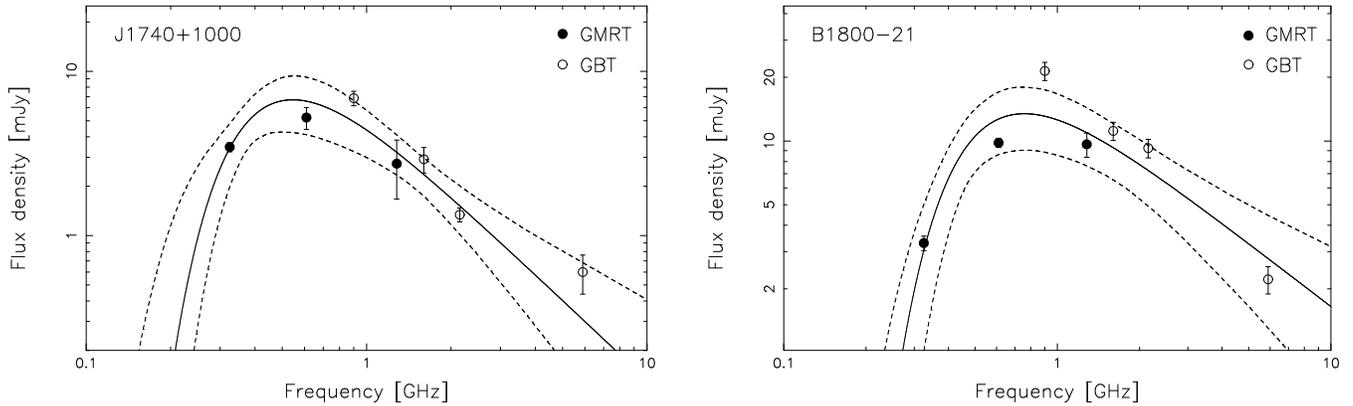}}
\caption{The pulsars spectra with fitted free-free thermal absorption model based on our latest flux density measurements. The dashed lines correspond to 1$\sigma$ envelope. The fitted parameters are presented in Table~\ref{tabRes1}.
\label{fits}}
\end{figure*}

As an additional check, we have measured the flux values of PSR~J1740$+$1000 and three other sources (the first two were the nearest ones and the third was the strongest source in the vicinity of the pulsar) breaking up each observing run at 1280~MHz into short 10-minutes scans.
This was done to identify the variations of the pulsar flux within a single session. Ten-minute sub-scans were the shortest possible, as we could not divide the data into shorter scans owing to the instability of the interferometric flux measurement technique at short durations. The results show that the flux density of PSR~J1740$+$1000 exhibit significant changes within each observing session. This was in sharp contrast to the surrounding sources which remained constant. Our result is consistent with the findings of \citet{mcl2002} which predict the diffractive scintillation time-scale at the 1410~MHz equal $\sim 5$ minutes. 

We believe that the variability seen in the flux measurements across different observing epochs is caused by scintillations. For this reason we believe that it is crucial to average the flux values over multiple sessions to obtain reliable measurements, especially at frequencies close to 1400~MHz. Our results clearly demonstrate the critical need for multi-frequency observations at multiple-epochs to establish the spectral properties and exclude biases due to scintillation related variability.

\subsection{The case of PSR~B1800$-$21 (J1803$-$2137)}
The flux density of PSR~B1800$-$21 for each frequency remains constant within the uncertainty level, only the  1280~MHz results show small fluctuations. The flux variations are likely caused by the refractive scintillation, which for this pulsar (at these frequencies) has a characteristic time-scale of days or weeks. Since we separated our observational epochs by a few weeks, this effect should not affect the mean flux density value. 

As can be seen in Figure~\ref{ap2} of the Appendix, this pulsar shows a significant profile evolution. To check if this may be the cause for the spectral turnover, we investigated the relative flux of the two main profile components. Obviously, any external cause to the observed turnover should affect both components in exactly the same way, while a strong profile evolution will cause the component intensities to change from frequency to frequency possibly affecting the total observed flux.

Since for the phased-array GMRT observations we only had uncalibrated profiles (the flux values in Table~\ref{p_ex} are based on interferometric observations) we could only check the relative flux ratio of the components. If we denote the flux of the left (lower) component as $S_{C1}$, and $S_{C2}$ is the flux of the higher component (see Figures~\ref{p_ex} and \ref{ap2}), then the lowest ratio of $S_{C2}/S_{C1}$ we measured was on the highest frequency (5900~MHz GBT observations) where $S_{C2}/S_{C1}=2.37$. This ratio is growing towards lower frequencies, although not monotonically, reaching the highest value of 4.79 at 325~MHz (however we have to note that this profile seems to be heavily affected by scattering). This clearly indicates a strong evolution, as the ratio changes by a factor of just over 2. 

A question we have to answer is if this kind of evolution may affect the shape of the pulsar spectrum and in what way. The observed profile evolution (see Figure~\ref{ap2}) may be caused by one of two possibilities: either the right component is getting stronger towards lower frequencies (its spectrum is steeper than the left component's), or the left component is getting weaker, its spectrum being less steep or even turning over. If the first scenario is true, then since the right component is dominating the total flux we would expect the spectrum of the pulsar to become steeper, and certainly not to exhibit a turnover at lower frequencies. If it is the left component that is getting weaker -- this should not affect the total flux by much, since this part of the profile contributes only 29\% of the total flux at the highest frequency ($S_{C1}/(S_{C1}+S_{C2})=1/(1+2.37)=0.29$) and only 17\% at the lowest. Clearly this would not be enough to explain the observed turnover, since the total pulsar flux at the lowest frequency (3.29~mJy) is an order of magnitude lower than the flux expected if the pulsar was exhibiting a simple power-law spectrum. Therefore, we can confidently say that the observed evolution of the profile shape does not affect the spectrum in any meaningful way and cannot explain the observed turnover.

Another way to confirm the external nature of the turnover would be to check the properties of individual pulses. Any external cause for the drop in the flux density should not affect the shape of the pulse-energy distribution for individual pulses; it should only change its characteristic flux value. There we encounter a significant caveat; one would be interested to check for this effect at the lowest frequencies, where the spectral turnover is strongest. Unfortunately, that means observations of single pulses at frequencies where the pulsar is weakest (and additionally, the Galactic background gets strongest), hence single pulse observations are extremely difficult. In the case of the pulsar B1800$-$21 (as well as J1740$+$1000) such observations are impossible using the instruments we employed. For PSR~J1740+1000 one would probably have to use the Arecibo telescope to perform such an analysis. In the case of PSR~B1800$-$21, due to its negative declination, there is no operational telescope that would allow us to reach the sensitivity required to observe individual pulses at the lowest frequencies.

Despite the fact that single pulse observations are currently unavailable, we believe that have sufficiently demonstrated that the profile evolution cannot explain the observed turnover in the spectrum. Based on our observations we claim that the turnover in the spectrum of PSR~B1800$-$21 is most probably due to an external cause and thermal absorption seems like the most plausible process.

\subsection{Thermal absorption model}
The free-free thermal absorption model was first used in order to explain turnover in pulsars spectra at low radio frequencies of 100~MHz by \citet{sie73} and was extended to explain the GPS nature of pulsars and magnetars by \citet{kijak2011b,kijak2013}.
Recently, it has been applied to model the GPS behaviour in more detail by \citealt{lewandowski2015a} and \citealt{rajwade2016}. In our approach we assumed that the intrinsic pulsar spectrum can be expressed by a power-law: $I_{\nu} = \mathrm{A} (\nu/\nu_{0})^{\alpha}$ and using an approximate formula for thermal free-free absorption (\citealt{RyLa79},\citealt{Wil2009}) we get the estimated flux ($S_{\nu}$) at any frequency ($\nu$) as:
\begin{equation}
 S_{\nu} = \mathrm{A} \left( \frac{\nu}{10} \right)^{\alpha} e^{-B\nu^{-2.1}} 
\end{equation}
where A is the pulsar intrinsic flux at 10 GHz, $\alpha$ is the pulsar intrinsic spectral index and $\nu$ is frequency in GHz. The parameter $B$ is defined as:
\begin{equation}
B = 0.08235 \times \left(\frac{T_{\mathrm{e}}}{\mathrm{K}} \right)^{-1.35}~\left(\frac{\mathrm{EM}}{\mathrm{pc}~\mathrm{cm}^{-6}}\right),
\end{equation}
where $T_{\mathrm{e}}$ is the electron temperature and EM is the emission measure.
 $A$, $\alpha$ and $B$ were free parameters in the fitting procedure. To fit the data we used the Levenberg-Marquardt non-linear least squares algorithm (\citealt{levenberg44}, \citealt{Marquardt63}) and estimated the errors using $\chi^2$ mapping \citep{press1992}. Table~\ref{tabRes1} shows the results of our fits ($\nu_{\mathrm{p}}$ is the peak frequency, i.e. a frequency at which the spectrum exhibits a maximum) and Figure~\ref{fits} shows the pulsar spectra with the fitted model with $1 \sigma$ envelopes.

Following \cite{basu2016} and \cite{kijak2017}, we used the pulsars' dispersion measure (DM) to constrain the electron density and temperature of the absorber. Similar to previous studies, we have assumed that half of the DM is contributed by the absorber. Using that assumption we calculated the EM for three absorber cases: a dense supernova remnant filament (with size equal $0.1$~pc), a pulsar wind nebula (with size equal to $1.0$~pc) and a warm HII region (with size equal $10.0$~pc). For each case the fitted value of parameter $B$ gave suitable constraints on the electron temperature. The results are listed in Table~\ref{tabRes2}.

\begin{table}
 \renewcommand*{\arraystretch}{1.5}
 \caption{Estimating the fitting parameters for the gigahertz-peaked spectra using the thermal absorption model.}
\begin{tabular}{c c c c c} \hline

 A & B & $\alpha$ & $\chi^2$ &  $\nu_{\mathrm{p}}$ \\ 
 & &  &  & GHz \\ \hline
 \multicolumn{5}{c}{J1740$+$1000}  \\
 $0.132^{+0.275}_{-0.094}$ & $0.22^{+0.11}_{-0.12}$ & $-1.61^{+0.66}_{-0.63}$ & $4.27$ & $0.55$ \\
  \multicolumn{5}{c}{B1800$-$21}  \\
 $1.65^{+1.52}_{-1.05}$ & $0.26^{+0.15}_{-0.10}$ & $-1.00^{+0.39}_{-0.49}$ & $8.94$ & $0.76$ \\
\hline
\end{tabular}
\label{tabRes1}
\end{table}

\begin{table}
 \renewcommand*{\arraystretch}{1.5}
 \caption{The constraints on the physical parameters of the absorbing medium.}
\begin{tabular}{c c c c} \hline
 Size & n$_{\mathrm{e}}$ & EM & T$_{\mathrm{e}}$ \\ 
 pc & cm$^{-3}$ & pc cm$^{-6}$& K\\ \hline
 \multicolumn{4}{c}{J1740$+$1000}  \\
 0.1 & $ 119.5\pm{0.1}$ & $1428\pm{3.0}$ & $106^{+42}_{-45}$ \\
 1.0 & $11.95\pm{0.01}$  & $142.8\pm{0.3}$ & $19.3^{+7.6}_{-8.2}$ \\
 10.0 & $1.195\pm{0.001}$ & $14.28\pm{0.03}$ & $3.5^{+1.4}_{-1.5}$ \\ \hline
  \multicolumn{4}{c}{B1800$-$21}  \\
 0.1 & $1170\pm{0.3}$ & $136880\pm{60}$ & $2680^{+1100}_{-770} $ \\
 1.0 & $117\pm{0.03}$ & $13688\pm{6}$ & $488^{+199}_{-140}$ \\
 10.0 & $11.7\pm{0.003} $ & $1368.8\pm{0.6}$ & $89^{+36}_{-25}$ \\ \hline
\end{tabular}
\label{tabRes2}
\end{table}
\section{Discussion}
In this section we discuss our results and we consider the possible absorbers for both pulsars. Additionally, in the case of PSR J1740$+$1000 we present the fit of the thermal absorption model to all available flux density measurements (see Figure~\ref{1740f2}). 

\subsection{The case of PSR~J1740$+$1000}
PSR~J1740$+$1000 is located at a relatively large distance from the galactic plane and has a very low DM~$=24$~pc~cm$^{−3}$ \citep{mcl2000}. This pulsar is also known to have a X-ray Pulsar Wind Nebulae (PWN) with a very extended tail (see \citealt{ka08}, \citealt{ka10}). As we mentioned in Section 3.1, the pulsar exhibits very strong diffractive scintillations at 1400~MHz that made the interpretation of its spectrum very difficult in the past. We want to point out that the previous spectral interpretations were based mostly on data that came from single epoch flux density measurements. In our latest observations for most frequencies we averaged the flux density over 2 or 3 observational sessions separated by a few weeks. Our results support the GPS interpretation of the pulsar's spectrum (see Figure~\ref{fits}). 
In such a scenario the most probable absorber is a partially ionized small molecular cloud along the line of sight of the pulsar. This is consistent with the estimated electron density value for the absorber thickness of 0.1~pc (i.e. 119~cm$^{-3}$). It is worth noting that the electron density in front of the shock in some bow-shock PWNe was estimated (from optical observations of atomic emission lines) to be of the order of $50-100$ cm$^{-3}$ (\citealt{h89}, \citealt{li05}), which also agrees with our estimates.

The physical constraints presented in Table~\ref{tabRes2} rule out an HII region as a possible absorber, because the related temperature is much too low for a typical HII region. On similar grounds, an absorption caused by the electrons located inside the bow-shock PWN is also doubtful. As we showed in our previous work \citep{lewandowski2015a} the free-free thermal absorption in the bow-shock PWNe is most efficient if we are observing the pulsar from behind the PWN, i.e. through its comet-shaped tail. In the case of PSR~J1740$+$1000 the observed PWN tail is very long (around 2~pc, see \citealt{ka08}), and a cometary shape can be easily discerned, which means that we are probably looking at the PWN from its side. The absorption may also happen in the partially ionized matter located around the ``head'' of the PWN. The pulsar's UV and X-ray emission can heat up and ionize the surrounding medium, giving rise to a small ionized region in the ISM (see e.g. \citealt{bl95} and \citealt{ke2001}). Moreover, around some known bow-shock PWNe an ionization pre-cursor was observed in the form of an H$\alpha$ recombination halo \citep[see ][]{brown2014}.

\begin{figure}
\resizebox{\hsize}{!}{\includegraphics{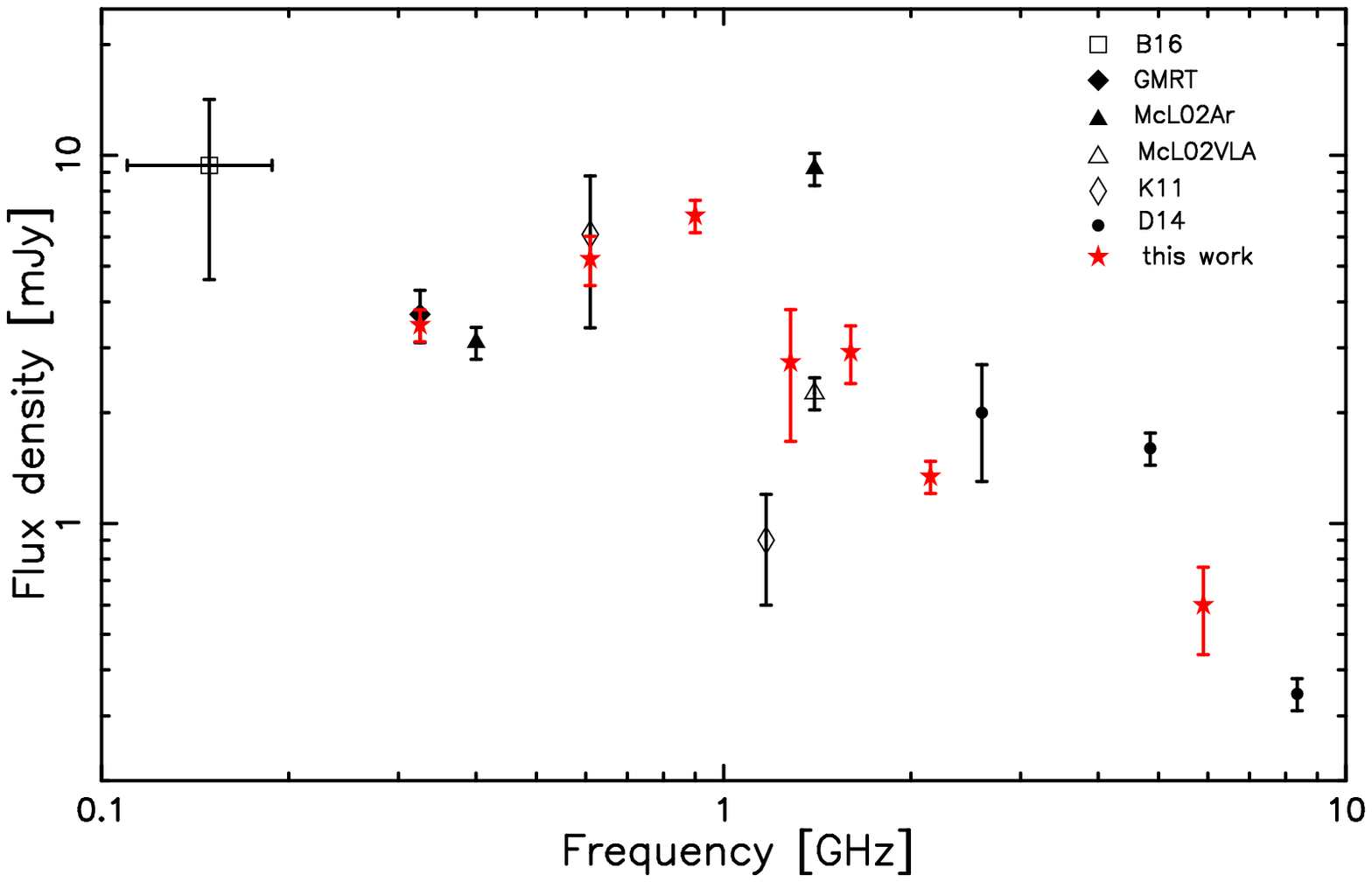}}
\caption{The figure shows the all flux density measurements for PSR~J1740$+$1000. The acronyms mean the following publications: B16 - \citet{bilous2016}, GMRT -  our interferometry measurements obtained using the GMRT that was not published previously (observed on January 2015), McL01Ar - \citet{mcl2002} using the Arecibo radio telescope, McL01VLA - \citet{mcl2002} using the Very Large Array, K11 - \citep{kijak2011b}, D14 - \citet{dembska2014}, red stars denote the flux density measurements from our recent observations. 
\label{1740f2}}
\end{figure}

\begin{figure}
\resizebox{\hsize}{!}{\includegraphics{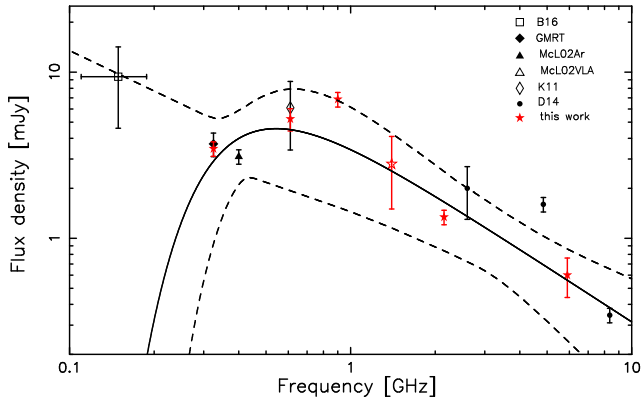}}
\caption{The PSR~J1740$+$1000 spectrum with fitted free-free thermal absorption model. The dashed lines correspond to 1$\sigma$ envelope. The open red star denotes the flux density value that was averaged over L-Band. The fitted parameters are presented in Table~\ref{1740tab}. The acronyms are the same as in Figure~\ref{1740f2}.
\label{1740f3}}
\end{figure}

As an additional exercise we applied the thermal absorption model to all the flux measurements available for this pulsar (including the 150~MHz measurements published by \citealt{bilous2016}), which are shown in Figure~\ref{1740f2}. We averaged all of the available measurements between 1180 and 1600~MHz to reduce the apparent spread which is due to the interstellar scintillations. As we mentioned in the section 3.1 the diffractive scintillation time-scale in this band is around 5-minutes. In the case of measurements based on short integrations, like the 12-minute integration made using the Very Large Array by \citet{mcl2002}, it is possible that the results obtained may be heavily affected by scintillations. For lower frequencies the diffractive scintillation time-scale decreases, hence the influence of the scintillation on flux density measurements will be weaker (since even a relatively short integration time will be enough to average-out the scintillation effect). For higher frequencies the diffractive scintillation time-scale will be even longer, however in our observations at frequencies above 2000~MHz we used the data from multiple observational epochs to minimize the influence of interstellar scintillations (with the only exception being the single measurement at 2150~MHz).
The best fit model still indicates a high frequency turnover, as seen in Figure~\ref{1740f3} and the model parameters are presented in Table~\ref{1740tab}. It should be noted that a single power-law spectrum also fits the data reasonably well (the $\chi^2$ for this model equals $11.64$), which can be seen from the of 1-$\sigma$ envelope, showed as the dashed lines in the plot. This is primarily caused by the 150~MHz LOFAR measurement from \cite{bilous2016}. Their measurement was based on a single 20-minute observation, and the profile upon which the measurement was based, has a low S/N ratio. In addition, the calibration procedure used for LOFAR observations results in 50\% errorbars \citep[see e.g.,][]{bilous2016,Kondratiev2016}. 

The large uncertainty of this measurement, combined with the modeling method we used (least squares fit with the data weighted by inverse square of the uncertainty) resulted in the near omission of the 150 MHz measurement. However, as one can see this point definitely affected the shape of the 1-sigma contour, and it is still possible to draw a single power-law spectrum that would in its entirety lie within 1-sigma of the best-fitting model we obtained.

Since the main goal of our study was to confirm the GPS character of the pulsar spectrum, and to test the hypothesis that it is thermal absorption that causes the turnover, we refrained from using more general methods (such as the robust techniques presented in~\citealt{hu1981}, and its application to the pulsar spectral analysis by \citealt{Jankowski2018}), since they are more useful for comparison between different spectral models. Our best fit clearly indicates a~GPS character of the spectrum, however a possibility of a single power-law can not be excluded.

At the moment lower frequency observations do not help to resolve this issue, since the only data that is available is the 25~MHz upper limit at 130~mJy \citep{zak13} does not allow us to draw any useful conclusions. 
The likely solution to resolving the low radio frequency spectral nature of PSR~J1740$+$1000 is accurate flux measurements around or below 200~MHz. Long observation using interferometric imaging would be the best way to carry out such studies

\begin{table}
 \renewcommand*{\arraystretch}{1.5}
 \caption{Estimating the fitting parameters for the gigahertz-peaked spectra using the thermal absorption model for all available measurements and averaged flux density over L-Band for PSR~J1740$+$1000}
\begin{tabular}{c c c c c c} \hline

 A & B & $\alpha$ & $\chi^2$ &  $\chi_{\mathrm{PL}}^2$ & $\nu_{\mathrm{p}}$ \\ 
  &  &  &  & & GHz \\ \hline
  $0.31^{+0.26}_{-0.21}$ & $0.15^{+0.21}_{-0.15}$ & $-1.10^{+0.56}_{-0.70}$ & $8.28$ & $11.64$ & $0.54$ \\
\hline
\end{tabular}
\label{1740tab}
\end{table}

\subsection{The case of PSR~B1800$-$21 (J1803$-$2137)}
Our observations confirm that the spectrum of PSR~B1800$-$21 is well described by a thermal absorption model with a peak frequency at 760~MHz. It means that the shape of the current spectrum is very similar to the spectral shape observed before 2009, when the peak frequency was equal 800~MHz. \citet{basu2016} showed that the peak frequency of the turnover in the spectrum shifted to a slightly higher frequency between 2012 and 2014. Our result would suggest that the spectral change observed during that period was indeed a rare event, rather than a continuous variation. The transition was explained by an additional absorption that appeared on top of a constant absorption component during that time. The most plausible explanation is that some small but dense supernova remnant filament crossed our line of sight. Since B1800$-$21 lies in the W30 complex, which is primarily a supernova remnant with a large number of HII regions scattered around it \citep{kw90}, such a scenario seems to be very plausible.

Apart from the additional absorption described in the previous paragraph, PSR~B1800$-$21 also exhibits an additional, constant absorption component. As showed by \citet{basu2016} the most plausible explanation is also the absorption in filamentary structures in the surrounding SNR. Our recent results confirm that conclusion and our estimate excluded an HII region as a possible absorber (see Table~\ref{tabRes2}). However, we cannot exclude the possibility that the GPS phenomenon arose in an asymmetric PWN.

\section{Conclusions}
We have conducted quasi-simultaneous, multi-frequency observations of two pulsars: J1740$+$1000 and B1800$-$21 using the GMRT and the GBT. Both pulsars were previously classified as GPS pulsars (see \citealt{dembska2014} and \citealt{dembska2015b}). 

In the case of the PSR~J1740$+$1000, the recent LOFAR measurements challenged the GPS interpretation of its spectrum \citep{bilous2016}. Our results suggest that the spectrum of this object exhibits a turnover at the frequency of 550~MHz. However, the low frequency behaviour of this pulsar spectrum still remains unclear. Only additional, independent measurements below $\sim 200$~MHz will allow us to obtain decisive conclusions. 

PSR~B1800$-$21 was reported to show spectral change on a~time-scale of a few years (see  \citealt{basu2016}). The analysis of our newest observations indicates that the spectrum of this pulsar is similar to that observed prior to the 2012 change (compare Figure~\ref{fits} and Figure 2 from \citealt{basu2016}). The most plausible interpretation for the  variation observed in the spectrum of this object is a relatively small filament crossing the line of sight.

Our analysis also shows that it is very important to have a wide frequency coverage when attempting to model gigahertz-peaked spectra. Flux measurements below the peak frequency allow us to estimate the amount of absorption, while high frequency measurements are crucial to ascertain the intrinsic pulsar spectrum. Finally, the case of PSR~J1740$+$1000 also highlights the problems with the estimation of the flux density at observing frequencies where the interstellar scintillations are significant. For this object the largest influence of the scintillations occurs at frequencies close to 1 GHz, i.e. close to the peak observed in the spectrum. This provides additional challenge when trying to describe the spectrum of PSR~J1740$+$1000. 

Our work shows that multi-epoch flux density measurements over a wide frequency range are crucial in order to accurately characterise the shape of pulsar spectra. Since the time-scale of the observed spectral changes (in the case of PSR B1800$-$21) is a few years we believe that the six months period of observations was sufficient to obtain reliable pulsar spectra. For the best results these observations should be at least quasi-simultaneous, i.e., separated by a few months at most. The spectral variability shown in the case of PSR B1800$-$21 happened over a few years, and one cannot realistically expect the time-scale of such changes to be much shorter.

Our analysis indicates that both pulsars show a turnover in the spectrum in the frequency range slightly below 1~GHz. However, especially in the case of J1740$+$1000, additional observations in the frequency range below 300~MHz may be necessary to confirm that free-free absorption alone is sufficient to explain the observed spectral features.

\section*{Acknowledgements}
We are very grateful to the anonymous referee for the comments that greatly improved
the paper. We thank the staff of the GMRT who have made these observations possible.
The GMRT is run by the National Centre for Radio Astrophysics of the Tata 
Institute of Fundamental Research. KMR thanks the NANOGrav PFC for providing flux calibrator observations to calibrate the GBT pulsar data. KMR acknowledges funding from the European Research Council grant under the European Union’s Horizon 2020 research and innovation programme (grant agreement No. 694745), during which part of this work was done. The National Radio Astronomy Observatory is a facility of the National Science Foundation operated under cooperative agreement by Associated Universities, Inc. This research was partially supported by the grant DEC-2013/09/B/ST9/02177 of the Polish National Science Centre.

\section*{Appendix}

In this appendix we present the integrated profiles obtained for PSRs~J1740+100 and B1800$-$21 from both the GBT observations as well as the GMRT phased array observations. The integration times were 45 minutes for the GMRT and 25 to 35 minutes for the GBT. Both pulsars show a clear evidence for the profile shape evolution, which in the case of PSR~B1800$-$21 is probably enhanced by the effect of interstellar scattering at the lowest observed frequency (325~MHz). PSR~J1740+1000 does not exhibit any signs of scattering in the frequency range we observed, which is understandable given its low DM value.

\begin{figure}
\resizebox{\hsize}{!}{\includegraphics{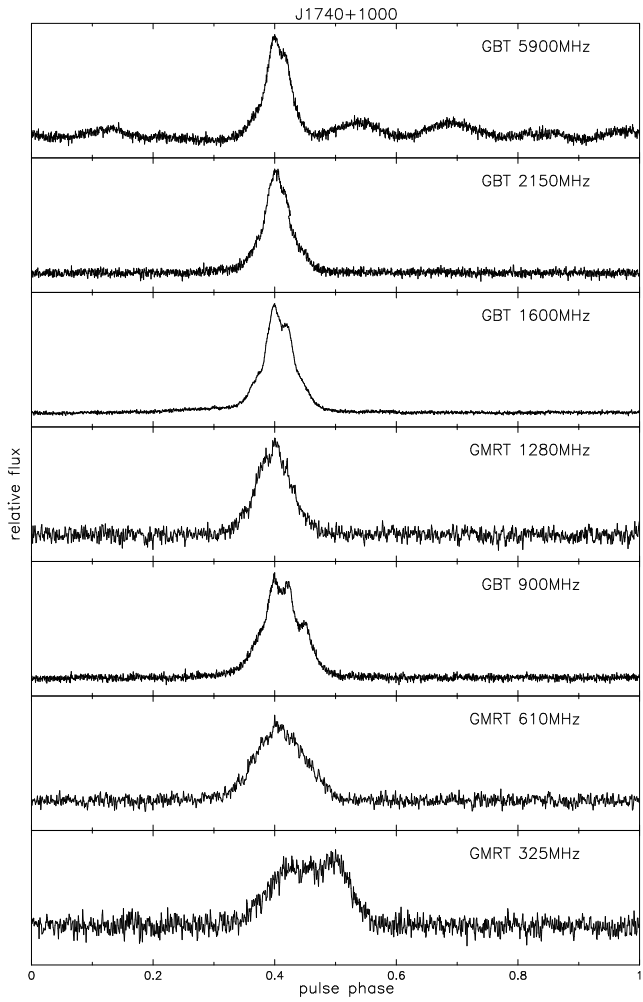}}
\caption{All profiles of PSR~J1740$+$1000.
\label{ap1}}
\end{figure}

\begin{figure}
\resizebox{\hsize}{!}{\includegraphics{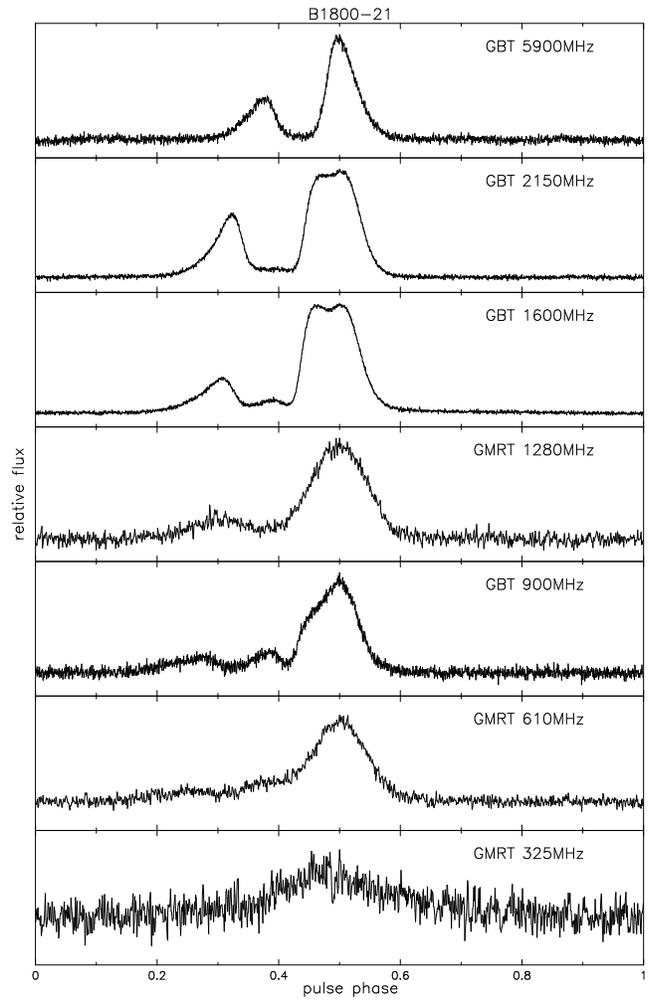}}
\caption{All profiles of PSR~B1800$-$21 (J1803$-$2137).
\label{ap2}}
\end{figure}

\end{document}